\documentclass[
reprint,
superscriptaddress,
showpacs,
preprintnumbers,
bibnotes,
amsmath,
amssymb, 
aps,
prl,
usenames,
dvipsnames,
floatfix
]{revtex4-2}

\usepackage{etoolbox} 
\newtoggle{PRL}
\togglefalse{PRL} 

\iftoggle{PRL}{
 \renewcommand{\section}[1]{}
 \renewcommand{\subsection}[1]{}
 \renewcommand{\subsubsection}[1]{}
 \renewcommand{\paragraph}[1]{}
 \renewcommand{\subparagraph}[1]{}
}{
}

\usepackage[utf8]{inputenc}
\usepackage[normalem]{ulem}
\usepackage{graphicx}
\usepackage{dcolumn}
\usepackage[colorlinks=true,allcolors=purple]{hyperref}
\usepackage{url}
\usepackage{hyperref}
\usepackage{enumerate}
\usepackage{accents}

\usepackage{xcolor}
\usepackage{slashed,multirow,relsize,soul,feynmp-auto,tikz}
\usepackage{mathrsfs} 
\usepackage{amsmath}
\usepackage{cancel}
\usepackage{bbold}
\usepackage{mathrsfs}
\usepackage{braket}
\usepackage{bm}
\usepackage{physics}
\usepackage{multirow}

\usepackage[capitalize]{cleveref}

\usepackage{xspace}
\usepackage{newunicodechar}
\newunicodechar{−}{-}
\usepackage{fontawesome} 
\definecolor{blue-violet}{rgb}{0.33, 0.17, 0.89}

\newcounter{CommentCount}
\definecolor{palatinate}{rgb}{0.494, 0.192, 0.482}
\definecolor{ferrarired}{rgb}{1.0, 0.11, 0.0}

\newcommand*{\pbar}[1]{\accentset{(-)}{#1}}

\newcommand{\thetaw}{{\theta_{\rm w}}}
\newcommand{\sinthetaw}{s_{\rm w}}
\newcommand{\Rthetaw}{R_{\rm w}}
\newcommand{\gitlink}{\href{https://github.com/mhostert/BIN_MC}{{\large\color{blue-violet}\faGithub} \textsc{g}it\textsc{h}ub}\xspace}

\begin{document}

\preprint{UMN-TH-4409/24}

\title{The Neutrino Slice at Muon Colliders}

\author{Luc Bojorquez-Lopez}
\email{luc\_bojorquezlopez@college.harvard.edu}
\affiliation{Department of Physics \& Laboratory for Particle Physics and Cosmology, Harvard University, Cambridge, MA 02138, USA}

\author{Matheus Hostert}
\email{mhostert@g.harvard.edu}
\affiliation{Department of Physics \& Laboratory for Particle Physics and Cosmology, Harvard University, Cambridge, MA 02138, USA}

\author{Carlos A. Argüelles}
\email{carguelles@g.harvard.edu}
\affiliation{Department of Physics \& Laboratory for Particle Physics and Cosmology, Harvard University, Cambridge, MA 02138, USA}

\author{Zhen Liu}
\email{zliuphys@umn.edu}
\affiliation{School of Physics and Astronomy, University of Minnesota, Minneapolis, MN 55455, USA}

\date{\today}

\begin{abstract}
 Muon colliders provide an exciting new direction to expand the energy frontier of particle physics.
 We point out a new use of these facilities for neutrino and beyond the Standard Model physics \emph{using their main detectors}.
 Muon decays along the accelerator rings create an intense and highly collimated neutrino beam that crosses a thin slice of the kt-scale detector.
 As a result, it would induce an unprecedented number of neutrino interactions, with $\mathcal{O}(10^4)$ events per second for a 10 TeV $\mu^+\mu^-$ collider.
 These interactions are highly energetic and possess a distinct timing signature and a large transverse displacement.
 We discuss promising applications of these events for instrumentation, electroweak, and beyond-the-Standard Model physics. 
 For instance, a sub-percent measurement of the neutrino-electron scattering rate enables new precision measurements of the Weak angle and a novel detection of the neutrino charge radius.
\end{abstract}

\maketitle


The discovery of the Higgs boson by the Large Hadron Collider (LHC) in 2012 was a major triumph for the Standard Model (SM) of particle physics~\cite{ATLAS:2012yve,CMS:2012qbp}. 
The new era centered around the fundamental puzzles of the SM such as the nature of electroweak symmetry breaking, the origin of neutrino masses, and the nature of dark matter now calls for exploration beyond TeV scales. 
With their compact size ($10$ km circumference), the high precision achievable in signal and background predictions, and the total number of Higgs bosons produced, muon colliders (MuC) with a center-of-mass energy of $\sqrt{s}\sim (1 - 10)$~TeV are strong contenders to push this energy frontier; their inherent complexity also presents new challenges and opportunities to develop new accelerator and detector technology.
A renewed interest in these facilities led to feasibility studies of $\mu^+\mu^-$ colliders by the International Muon Collider Collaboration (IMCC)~\cite{IMCCwebsite,Accettura:2023ked,InternationalMuonCollider:2024jyv} and the US muon collider community~\cite{AlAli:2021let,Black:2022cth}, efforts that are informed by the European Strategy for Particle Physics~\cite{Adolphsen:2022ibf} and the P5 report in the US~\cite{Narain:2022qud,MuonCollider:2022nsa,P5:2023wyd} as well as by past work by the Muon Accelerator Program (MAP)~\cite{MAPwebsite} (see also~\cite{Gallardo:1996aa,Ankenbrandt:1999cta}).
New ideas for $\mu^+$ beams based on muonium cooling are also being considered at J-PARC~\cite{Kondo:2018rzx,Hamada:2022mua}.

\begin{figure}[t]
\centering
\includegraphics[width=0.49\textwidth]{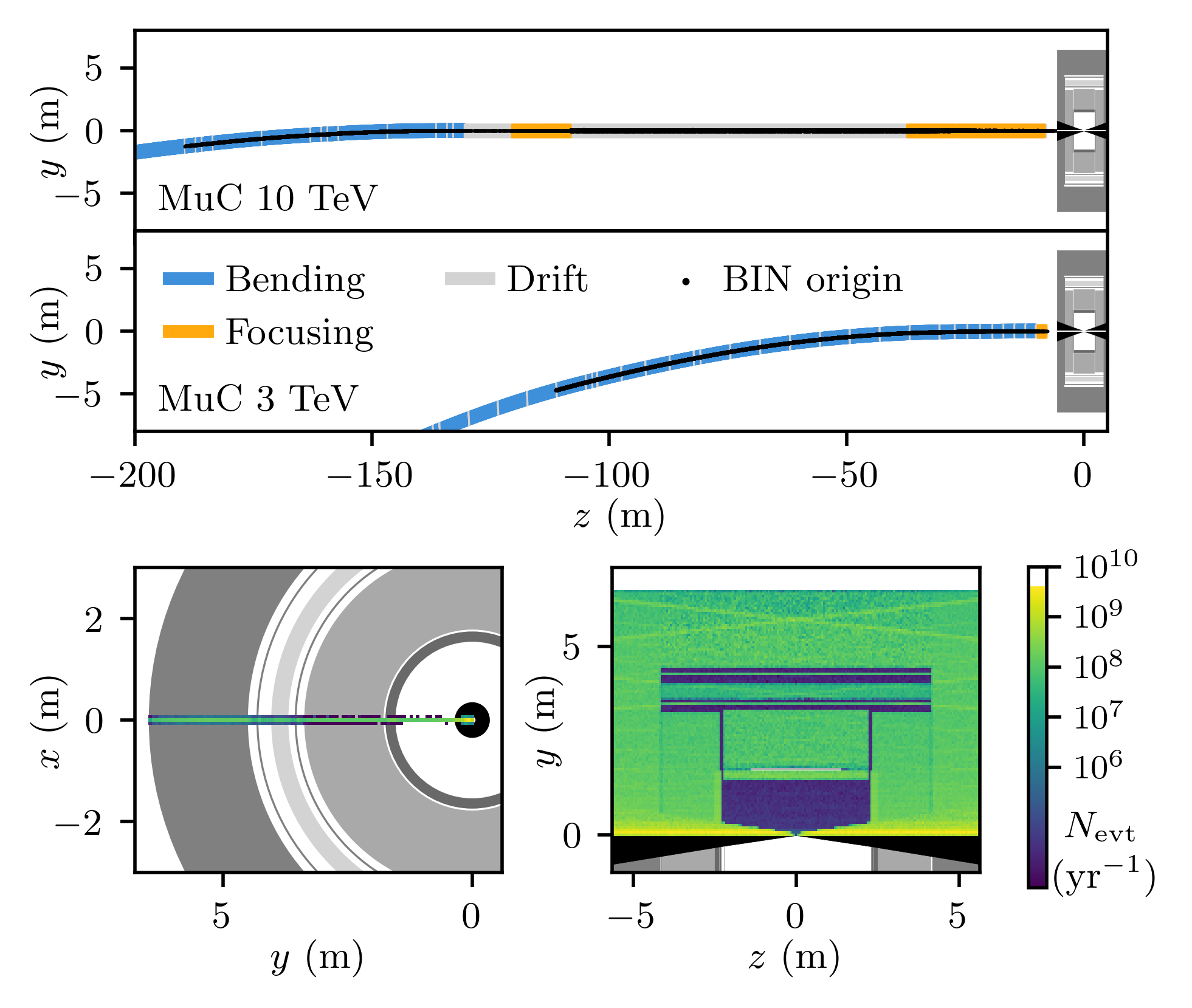}
\caption{Top: the positions of the muon decay (color) along the ring with a BIN crossing the main detector (gray region) for the two benchmark MuC designs, highlighting the different magnet systems. 
Bottom: a radial (left; the \emph{neutrino slice}) and top-down (right) view of the detector barrel and the number of BIN interactions from both muon beams in a year of operation of the MuC 10 TeV design.
Black regions indicate the tungsten nozzles.
\label{fig:diagram}}
\end{figure}

Even if low-emittance muon beams are demonstrated, a MuC design still requires optimization of the detector and muon interaction region (IR) to minimize the impact of beam-induced backgrounds (BIB) coming from $\mu^\pm \to e^\pm\nu\overline\nu (\gamma) $ decays and their secondaries.
An accelerated bunch of $N_\mu$ muons decays at a rate of
$R = (N_\mu m_\mu)/(p_\mu \tau_\mu) \sim 1.6 \times 10^{5} \text{ m}^{-1} \left({N_\mu}/{10^{12}}\right)\left(p_\mu/{1\text{ TeV}}\right)^{-1}$,
where $p_\mu$ is the momentum of the muons and $\tau_{\mu} = 2.2 \,\mu$s, $m_{\mu}$ the muon lifetime and mass, respectively.
While the accelerator magnets deflect charged particles, neutrinos will travel tangentially to the beam and cannot be shielded against; as a result, beam-induced neutrinos (BIN) will produce highly energetic muons, electrons, and hadrons through Weak interactions with detector materials.
These BIN interactions would populate the entire radius of the barrel-shaped main detector at the IR, spanning a small azimuthal slice aligned with the plane of the collider ring (hereafter referred to as the \emph{neutrino slice}, see, ~\cref{fig:diagram}).
This behavior of BINs is in stark contrast with ordinary BIBs, which mostly affect the inner layers of the detector, beam pipes, and magnets in almost cylindrically symmetric ways.
BINs originate from as far as $\mathcal{O}(30-100)$~m from the IR and can be asynchronous to the bunch crossings.

This \emph{letter} represents a first step in exploring the BIN physics program at a MuC.
With current targets for muon beam power at future MuCs, BINs would provide the \emph{largest sample size} of detectable neutrino-nucleus and neutrino-electron interactions and the \emph{highest-energy}, \emph{most collimated}, and \emph{most well-characterized} neutrino beam ever produced in a laboratory.
To exploit this unique neutrino beam, rejecting backgrounds from beam loss and muon-decay BIBs is crucial; this calls for future studies that leverage the distinct energy and spatiotemporal profile of BINs.
Detection of the most forward part of BINs is also complementary to new ideas to tag forward-going muons to measure Higgs boson properties~\cite{Forslund:2022xjq,Ruhdorfer:2023uea,Li:2024joa,Forslund:2023reu,Ruhdorfer:2024dgz} and the Higgs boson width~\cite{Li:2024joa}.
While BINs have been studied before as a radiation hazard on the surface of the Earth~\cite{King:1999ja,King:1999kx,Bevelacqua:2012odt,MuonCollider:2022ded} and for forward detectors~\cite{King:1997dx}, to the best of our knowledge, this is the first study of tangential BIN interactions in a MuC detector.

Unlike a neutrino factory (a racetrack-shaped ring design to store $\mathcal{O}(10-20)$~GeV~\cite{Geer:1997iz,Mangano:2001mj,IDS-NF:2011swj,Bogomilov:2013zba,Bogacz:2022xsj} or $\mathcal{O}(1 - 6)$~GeV~\cite{nuSTORM:2012jbd} muons), a MuC would not be well suited for oscillation studies within the SM due to the higher energies involved. Instead, BINs offer the opportunity to study fundamental physics through precision measurements of the Weak force with neutrinos, overcoming flux systematics that plague conventional neutrino beams.
As such, a tangential BIN physics program using the main MuC detector could complement forward physics facilities at a MuC \textit{\`a la} FASER at the LHC~\cite{FASER:2019dxq,FASER:2020gpr,FASER:2021mtu}.

\emph{Beam induced neutrinos---}To evaluate the number of BIN interactions at a MuC we developed a Monte Carlo simulation of muon decays around the main collider ring.
The muons are modeled according to the average behavior of the beam in the magnet lattice designs provided by the IMCC~\cite{chance_2024_13847233,calzolari_2024_14000854,MuCoL:2024oxj}, described in more detail in the \emph{Supplemental Material}.
The beam size and angular aperture are approximated by Gaussians with width determined by the Twiss parameters in the lattice.
We assume a fixed number of muons per bunch $N_{\mu}$ for each collider, with an injection rate of $f_{\rm inj}$ and bunch multiplicity of $N_{\rm bunch}$.
The number of bunch crossings per second is then approximately $f_{\rm inj} N_{\rm bunch} n_{\rm turns}$, where $n_{\rm turns} \simeq (\tau_\mu \gamma_\mu \beta_\mu c)/C$ is the typical number of turns around the ring of circumference $C$.
The beam momentum spread is assumed to be a constant Gaussian distribution centered around $p_{\rm beam} \simeq \sqrt{s}/2$  with $\sigma_p/p_{\rm beam} = 0.1\%$.
The spatial and angular spreads of the muons are modeled as Gaussians that follow the beam envelopes in the transverse directions.\footnote{
Our setup captures only the average behavior of the muon trajectories as these oscillate throughout the ring.
A full beam simulation that tracks individual muons throughout the lattice will enable more accurate spatial and temporal BIN distributions than what is done in this first study.}

We consider three benchmark designs:
\begin{itemize}
\item{MuC 3 \& 10 TeV}:
we choose the $\sqrt{s} = 3$ and $10$ TeV designs defined by current muon collider benchmark studies~\cite{InternationalMuonCollider:2024jyv,Black:2022cth}.
More details on the IP and machine designs can be found in \cite{Bartosik:2019dzq,Collamati:2021sbv,MuonCollider:2022ded,Accettura:2023ked,MuCoL:2024oxj,Bartosik:2024ulr}.
As we shall see, the total BIN interaction rate for these designs are $2.0\times 10^{10}$ and $1.5 \times 10^{11}$ events per year, respectively.

\item{$\mu$TRISTAN 2 TeV}: a collider inspired by the muonium cooling method for $\mu^+$ at J-PARC~\cite{Abe:2019thb,Hamada:2022mua}.
While both $\mu^+e^-$ or $\mu^+\mu^+$ colliders are possible, we consider the latter where each beam has $E_\mu = 1$~TeV.
Note that the bunch injection frequency and the number of bunches per cycle assumed are an order of magnitude higher than in the designs above.
We also assume a polarization of $\mathcal{P}_{\mu^+} = 0.8$ can be achieved~\cite{Hamada:2022mua}.
For this design and our detector choice, we find $1.6\times 10^{11}$ BIN interactions per year.
\end{itemize}

We consider a single barrel-shaped detector design for all MuC benchmarks.
It emulates the detector model for CLIC~\cite{CLICdp:2017vju} with the addition of two conical tungsten nozzles to shield active detectors against BIBs~\cite{MuonCollider:2022ded}.
The shielding was optimized for the MAP $\sqrt{s} = 1.5$~TeV collider~\cite{MuonCollider:2022nsa}, so while we expect the design to evolve on the road to a 10 TeV MuC, some of its features are general.
Notably, the detector consists of thin silicon layers for the tracking system enveloped by electromagnetic and hadronic calorimeters which, in turn, are surrounded by the solenoid and muon detector layers.
All materials and dimensions of individual components are detailed in the \emph{Supplemental Material}. 
In this work, we neglect the thin tracker layers, assume a vacuum in the beampipe, and set the density of the remaining detector regions to that of air.

\cref{fig:diagram} shows the general shape of the BIN interaction region within the detector and \cref{tab:event_rate_makeup} quantifies the interaction rates of our three benchmark scenarios. 
We characterize the kinematics and geometry of the events in greater detail in the Endmatter.
The BIN event rate in the entire kiloton-scale detector is of the order of $(0.005 - 0.5)$ per bunch crossing, resulting in as many as $10^{11}$ BIN interactions/year within the nozzles, the calorimeters, and the muon detector. 
This event rate corresponds to about $1\times10^{5}$ BIN interactions per Higgs boson produced at a 10 TeV MuC for an assumed luminosity of $1$~ab$^{-1}$/year; for comparison, the Deep Underground Neutrino Experiment (DUNE)~\cite{DUNE:2021nd,DUNE:2021tad} expects about $10^7 - 10^8$ events per year at a 50~t near detector.

We now consider the feasibility of measuring BIN interactions in a MuC detector.
A precision measurement program will rely on low background and good reconstruction levels; for the latter, there are four main event categories to consider:
i) Neutral Current (NC) Deep Inelastic Scattering (DIS),
ii) $\nu_e$ Charged Current (CC) DIS,
iii) $\nu_\mu$CC DIS,
and iv) fully leptonic, coherent, or low-hadronic-energy events.
The first three categories contain an energetic hadronic shower and are best studied in the HCAL and ECAL.
For (iii), the additional muon can be independently identified and its transverse momentum can be measured with the magnetic field in the detector.
We note that, as per \cref{tab:event_rate_makeup}, 50 - 80 $\%$ of BIN interactions will occur in the BIB-shielding nozzles and beam pipe and therefore cannot be precisely reconstructed; however, a large number of BINs (mainly originating from the final arcs before the IP straight section) will intersect with the HCAL, ECAL, and muon detectors, subtending an angle of $0.1^\circ-2.5^\circ$ ($0.6^\circ - 6^\circ$) with respect to the detector axial direction at the 10 TeV (3 TeV) MuC.
The reconstruction of such shallow-angle particles will greatly benefit from a longitudinally segmented detector.

The electrons and muons produced in BIN interactions in the ECAL and HCAL carry a typical transverse momentum of $\langle p_T^\ell \rangle \simeq 37$~GeV for MuC-10-TeV, $32$~GeV for MuC-3-TeV, and $21$~GeV for $\mu$TRISTAN.
The typical travel distance of electrons and muons in the radial direction of the ECAL and HCAL detectors is about $10$~cm, $27$~cm, and $27$~cm, respectively.
Assuming calorimeters similar to that of the HGCAL at CMS~\cite{Liu:2020vur,CERN-LHCC-2017-023}, it is conceivable that these events would be reconstructed with good energy and spatiotemporal resolution.
Nevertheless, we urge detector R\&D studies to consider finer granularity in the radial and axial direction that subtends the \emph{neutrino slice} of \cref{fig:diagram}.

The BIB occupancy in the ECAL and HCAL will determine the energy resolution and threshold for BIN studies.
Thanks to the tungsten nozzles around the IR, hadrons and charged particles lose most of their energy by the time they enter the active components, making BIBs with energy above $1$~GeV extremely unlikely~\cite{Mokhov:2011zzd,Collamati:2021sbv}.
This is even more true in the calorimeters located much further from the IR than the tracking system. 
Still, within the first nanoseconds of the bunch crossing, BIB occupancy impacts energy resolution to single $\mathcal{O}(100)$~MeV energy depositions from BINs.
This effect might also limit the minimum hadronic energy required to distinguish BIN interactions with and without hadronic showers.

\begin{figure}[t]
  \centering
  \includegraphics[width=0.49\textwidth]{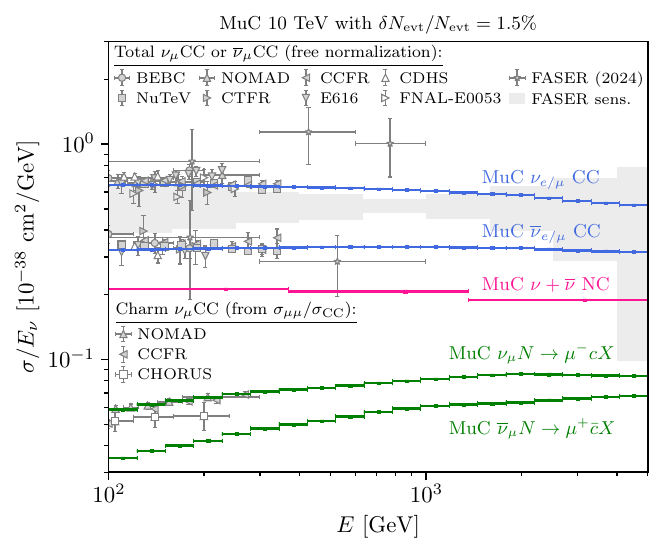}
  \caption{
 The inclusive neutrino cross section as would be measured at the 10 TeV MuC as a function of energy for $\nu_{e/\mu}$ and $\overline{\nu}_{e/\mu}$.
 Measurements of the $\nu_\mu$CC and $\overline\nu_\mu$CC inclusive cross sections are shown as data points. 
 In most cases, the measurements were normalized to global neutrino data to circumvent flux uncertainties.
 We also show data on charm production by normalizing the dimuon cross section $\sigma_{\mu\mu}$ by the $c \to \mu$ branching ratio reported by NOMAD~\cite{NOMAD:2013hbk}.
 In all cases, the underlying theory predictions are from \cite{Weigel:2024gzh}.
 }
  \label{fig:xsec_measurement}
\end{figure}

\emph{Backgrounds ---}We now comment on potential sources of backgrounds to BINs, highlighting the need for future investigation by machine design studies.
Products of $\mu^+\mu^-$ collisions are relatively rare and identifiable from the IP, so backgrounds to consider will arise from the beam itself.
There are key distinguishing features between muon-decay BIBs and BIN interactions.
Firstly, BIN final states carry a significant fraction of the beam energy, resulting in TeV-scale leptons and hadrons at \textit{large detector radii}; this is in stark contrast with muon-decay BIB (non-BIN) particles which come in high multiplicities but are more diffuse and less energetic. 
Requiring a starting event (resolvable interaction vertex within the ECAL or HCAL) will be crucial to reconstructing BIN events.
Secondly, the spatio-temporal distribution of muon-decay BIBs and BINs are very different: BIBs are distributed evenly across the detector, with an approximate $\phi$ symmetry (in stark contrast with the neutrino slice) and their time of detection is highly asymmetric about the bunch collision, starting about 20 ns before and lasting at least 100 ns after, sharply peaked at the bunch crossing~\cite{Bartosik:2024ulr}.
On the other hand, the BIN time distribution is flat around the bunch crossing spanning an interval of $\pm 20$~ns (for calorimeter events, this interval is closer to $\pm 14$~ns).
This is longer than the typical $\mathcal{O}(30 - 60)$~ps time span of the muon beam-buckets and much shorter than the bunch revolution time of $\mathcal{O}(30)$~$\mu$s.

Another important source of BIBs comes from beam halo losses due to large-amplitude betatron oscillations, residual gas interactions, and beam-beam effects. 
These halo muons escape the bucket, penetrate the shielding, and reach the neutrino slice. 
For earlier designs~\cite{Gallardo:1996aa}, a loss rate of $10^{-6}$ per muon-turn was targeted. 
Assuming uniform tangential losses along the ring, the MuC 10 TeV design would yield $\mathcal{O}(10)$ muons within detector acceptance per bunch crossing. 
We note, however, that loss rates remain unstudied for modern benchmarks~\cite{InternationalMuonCollider:2024jyv,MuCoL:2024oxj}. 
These muons, though more angularly dispersed, still point toward the neutrino slice and have energies just below the beam energy. 
Deflection using beam extraction and scraping near the IP may reduce this~\cite{Drozhdin:1998yg}. 
Undetected muons, including those from Bethe-Heitler processes, can undergo catastrophic energy losses in calorimeters, mimicking EM final states. 
Mitigation may come from fiducialization and efficient vetoing using, e.g., the outer muon spectrometer endcap. 
While BINs, decay BIBs, and halo muons differ, addressing them requires R\&D on the neutrino slice and studies of halo scraping/sweeping, detector timing, and longitudinal segmentation.

\emph{Flux uncertainty---}The uncertainty in the BIN flux will be dominated by the muon beam uncertainties.
It can be reduced in two main ways that are independent of the BIN measurement~\cite{ISSDetectorWorkingGroup:2007uvu}: (i) with beam current transformers, subtracting parasitic currents, and (ii) with beam monitors that detect deflected electron and positrons in specific momentum and spatial bites.
Option (i) was studied for a low-energy neutrino factory, where it was shown that the parasitic currents are safely below the $10^{-3}$ level~\cite{Keil:2000tq,Zimmermann:702631}.
At the LHC, the current precision is at the $1\%$ level~\cite{Odier:1372191} or better~\cite{Krupa:2023mhg}.
Option (ii) would also be valuable, especially at the arcs before the long straight section where most BINs will come from.
This can also allow a direct measurement of the beam polarization~\cite{Raja:1998ip}.

Considering both the high collimation and the gargantuan statistics, the uncertainty on the predicted BIN rate will likely be dominated by systematic errors on efficiencies, backgrounds, and the number density of scatterers inside the detector.
Furthermore, theoretically-clean BIN events like neutrino-electron elastic scattering can provide a statistical measurement of the product of BIN flux and electron density.

\begin{figure}[t]
 \centering
 \includegraphics[width=\linewidth]{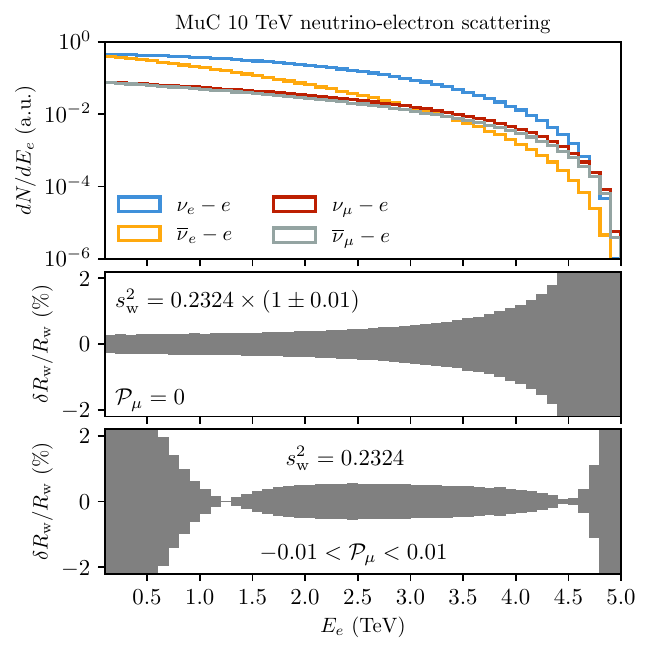}
 \caption{Top) The central value of the electron spectrum of ES events for different neutrino flavors at the MuC 10 TeV setting $\sinthetaw^2 = 0.2324$ and $\mathcal{P}_\mu = 0$.
 Middle) the relative change in the ratio $\Rthetaw$ (\cref{eq:Rthetaw}) by varying $\sinthetaw^2$ by $1\%$.
 Bottom) the relative change when varying the beam polarization by $1\%$.
 \label{fig:polarization_ES}
 }
\end{figure}

\emph{Neutrino cross sections---} 
The large flux of energetic BINs at muon colliders allows a direct precision measurement of inclusive and exclusive neutrino-nucleus and neutrino-electron cross sections in the TeV range.
With sufficient background control, one can achieve subpercent-level precision, an order of magnitude improvement over conventional accelerator neutrino experiments.
Indeed, overcoming flux uncertainties is the hallmark of neutrino factories (and of BINs at a MuC) and would enable a rich leptonic and hadronic physics program (see, e.g.,~\cite{Mangano:2001mj,IDS-NF:2011swj,Bogomilov:2013zba,Bogacz:2022xsj}).
Neutrino-nucleus deep inelastic scattering would span a broad range of $Q^2$ and $x$ values, bringing the ongoing program to extract nuclear parton-distribution functions using neutrinos at the LHC~\cite{FASER:2020gpr,Cruz-Martinez:2023sdv,Gao:2017yyd} to the level of sub-percent precision.

\cref{fig:xsec_measurement} shows the potential of a 10 TeV MuC in measuring neutrino DIS cross sections under a conservative assumption of $1.5\%$ normalization uncertainty.
We neglect nuclear effects, which can lead to a reduction of the rate shown by as much as $\mathcal{O}(10)\%$~\cite{Klein:2020nuk}.
Neutrino energy reconstruction is more challenging for NC events, so we adopt a conservative binning in $E_\nu$ (see \cite{Ismail:2020yqc} for a discussion of NC events at FASER).
Other accelerator experiments are shown as gray datapoints, but we note that their overall normalization was often fit to global neutrino data (see \cite{Conrad:1997ne} for a review).
The latest measurements and projected sensitivity of FASER~\cite{FASER:2023zcr,FASER:2024ref}, shown as large error bars and as a grey band, respectively, are limited by hadronic uncertainties in the neutrino flux prediction.
We also show how a $1.5\%$ extraction of the charm-production cross section in both $\nu_\mu$ and $\overline{\nu}_\mu$ beams compares to previous measurements~\cite{NuTeV:2001dfo,NuTeV:2007uwm,CHARMII:1998njb,CHORUS:2008vjb,CCFR:1994ikl,NOMAD:2013hbk}.
This channel can be accessed with dimuon final states and constrains the strange content of the nucleon, the charm quark mass, and the CKM parameter $|V_{cd}|$.
Note that the total cross section is only reported by a few experiments.

\emph{Electron Scattering (ES)---} 
Neutrino scattering on electrons is sensitive to electroweak parameters like $\thetaw$ and neutrino electromagnetic properties at intermediate momentum transfers ($30$~MeV $\lesssim Q \lesssim 2$~GeV)~\cite{deGouvea:2006hfo,Kouzakov:2017hbc,deGouvea:2019wav,MammenAbraham:2023psg}.
The signal is a forward-going electron shower without any hadronic activity, respecting $E_e \theta_e^2 < 2 m_e$, with $E_e$ and $\theta_e$ the electron's energy and angle with respect to the incoming neutrino.
Each location in the detector corresponds to a given tangential location on the ring, so $\theta_e$ can be reconstructed on an event-by-event basis within a given resolution.
With an unprecedented signal sample size, this measurement will be limited by systematic uncertainties, many of which can be reduced by exploiting the unique properties of BINs.
To illustrate this point, let us consider a z-symmetric detector and identical $\mu^+$ and $\mu^-$ beams.
Neglecting the difference in shape of the $\nu_e$ and $\nu_\mu$ fluxes, the relative difference of ES from the $\mu^+$ and $\mu^-$ beams is
\begin{align}\label{eq:Rthetaw}
\Rthetaw &= \frac{N_{\mu^+} - N_{\mu^-}}{N_{\mu^+} + N_{\mu^-}} =\frac{ N_{\overline\nu_\mu} + N_{\nu_e} - N_{\nu_\mu} - N_{\overline\nu_e}}{N_{\overline\nu_\mu} + N_{\nu_e} + N_{\nu_\mu} + N_{\overline\nu_e}} 
\\ \nonumber
&\simeq 2 \sinthetaw^2 \left(\frac{1 + r}{
  1 + 4 \sinthetaw^2 \delta r + 8 \sinthetaw^4 (1 + 2 r + r^2 + \delta r^2) 
  }\right),
\end{align}
where $r = \frac{m_W^2}{3} (\langle r_{\nu_e}^2\rangle + \langle r_{\nu_\mu}^2\rangle)$ and $\delta r = \frac{m_W^2}{3} (\langle r_{\nu_e}^2\rangle - \langle r_{\nu_\mu}^2\rangle)$ parameterize the neutrino charge radius.
The precision on the extraction of $\sinthetaw$ or on the charge radius, for instance, is then dictated by the relative difference of events towards and away from the beam axis (left- and right-going electrons) in a MuC detector.
The $z$-symmetric backgrounds and normalization systematics cancel in the ratio in an ideal geometry.
Note that the four elastic scattering channels ($\bar\nu_\mu,\nu_e,\nu_\mu,\bar\nu_e$) and neutrino fluxes have different and well-known dependencies in energy.
A full fit to the energy, angle, vertex location, and timing of electron events will further boost the sensitivity to physical parameters.

\cref{fig:polarization_ES} shows the electron recoil energy $E_e$ dependence of $\Rthetaw$, accounting for the full energy dependence on neutrino fluxes.
We also show the variation in $\Rthetaw$ with $\sinthetaw^2$ and the muon polarization $\mathcal{P_\mu}$.
A percent measurement of $\Rthetaw$ clearly enables the 10 TeV MuC BIN events to improve on previous neutrino measurements of $\sinthetaw^2$.
The ES measurement at CHARM-II~\cite{CHARM-II:1994dzw} gives $\sinthetaw^2 = 0.2324 \pm 0.0083$, which, neglecting the energy dependence, gives $\Rthetaw = (0.3244 \pm 0.0046)$ ($\delta \Rthetaw/\Rthetaw < 1.4\%$).
Generally, however, atomic parity violation measurements are more sensitive probes of $\sinthetaw$ thanks to the interference between the $Z$ and photon diagrams.
A $\delta \Rthetaw/\Rthetaw$ precision better than $0.20\%$ and $0.06\%$ is needed to beat the latest measurement by Qweak ($\sinthetaw^2 = 0.2383 \pm 0.0011$)~\cite{Qweak:2018tjf} and future projects~\cite{Demiroglu:2024wys}, respectively.
Using such measurements as an input, one can then extract for the first time the neutrino charge radius and search for new physics.
In the SM, $\langle r_{\nu_e}^2\rangle = 4.1\times 10^{-33}$~cm$^2$ and $\langle r_{\nu_\mu}^2\rangle = 2.4\times 10^{-33}$~cm$^2$~\cite{Bernabeu:2000hf,Bernabeu:2002pd}, so for the Qweak $\sinthetaw$ value, a precision of $\delta \Rthetaw/\Rthetaw = 0.17\%$ is enough for a first detection of the $\nu_e$ and $\nu_\mu$ charge radii at the $95\%$~CL.

These are just some of the applications of a precision measurement of BINs at a MuC. 
In the Appendix, we present greater details of exclusive neutrino scattering channels, potential effects of muon polarization, as well as our simulation prediction for the forward neutrino fluxes.
We also note that several new physics searches discussed for the forward direction~\cite{King:1997dx,Bigi:2001xb,Adhikary:2024tvl,Liu:2024ywd} and beam dumps at a MuC~\cite{Cesarotti:2022ttv,Cesarotti:2023sje} could be adapted for the neutrino slice.

\emph{Summary---}BIN interactions in the MuC detector are so frequent that they i) would bring about an unavoidable physics program to MuCs and ii) should be considered as part of ongoing machine design efforts, especially in view of its complementarity with BIB-mitigation efforts.
In particular, the concentration of these events in the \emph{neutrino slice} invites new detector concepts that break the $\phi$ symmetry of the detector, e.g., a new high-density and high-granularity slice to further enhance the neutrino program.
Accelerator and detector design for future MuCs is a rapidly developing research topic.
For example, alternative detector designs such as the MUSIC and MAIA concepts~\cite{MuCoL:2024oxj} move the solenoid inwards before the HCAL and ECAL, respectively, decreasing the BIB rate in the calorimeters and potentially enhancing the BIN reconstruction capabilities.
Moving forward, a full detector simulation of BINs is in order to more accurately quantify the sensitivity of BINs as beam monitors and electroweak and new physics precision probes.
With this letter, we hope to encourage new MuC R\&D to optimize the physics potential of this unique beam of neutrinos.

\emph{Acknowledgments---}
We thank Daniele Calzolari and Rodolfo Capdevilla for discussions on the beam-induced neutrino flux, Ryuichiro Kitano for suggesting the use of muon polarization, and André de Gouvea and Adrian Thompson for discussions on neutrino-electron scattering.
LBL was supported by the \textit{Harvard College Research Program} (HCRP) and funding from the Faculty of Arts and Sciences of Harvard University through this work.
The work of MH is supported by the Neutrino Theory Network Program Grant \#DE-AC02-07CHI11359 and the US DOE Award \#DE-SC0020250.
ZL is supported by the DOE grant \#DE-SC0011842 and a Sloan Research Fellowship from the Alfred P. Sloan Foundation at the
University of Minnesota.
CAA are supported by the Faculty of Arts and Sciences of Harvard University, the National Science Foundation (NSF), the NSF AI Institute for Artificial Intelligence and Fundamental Interactions, the Research Corporation for Science Advancement, and the David \& Lucile Packard Foundation.

The code used for this research is made publicly available through \gitlink under CC-BY-NC-SA.

\appendix
\section*{Endmatter}

\subsection{Characterizing BIN events}
\label{app:characterizingBINs}


\begin{table}[t]
  \centering
  \renewcommand{\arraystretch}{1.25}
  \begin{ruledtabular}
  \begin{tabular}{lccc}
     Collider 
     & MuC 10 TeV & MuC 3 TeV & $\mu$TRISTAN 
    \\
    \hline
        Beams 
        & $\mu^+ \mu^-$& $\mu^+ \mu^-$& $\mu^+ \mu^+$
    \\  
        Muons/bunch 
        & $1.8\times 10^{12}$ & $1.8\times 10^{12}$ & $1.4\times 10^{10}$ 
    \\  
        bunches/cycle
        & 1 & 1 & 40 
    \\  
        $f_{\rm inj}$
        & 5~Hz& 5~Hz& 50~Hz
    \\  
        $C$
        & 8.7~km& 4.3~km& 4.3~km
    \\  
    \hline\hline
    \multicolumn4{c}{BIN inclusive reactions}
    \\
    \hline
    ECal ($0.15$~kt)
    & 0.9\% & 3.0\% & 3.0\% 
    \\
    HCal ($1.4$~kt)
    & 7\% & 15\% & 15\% 
    \\
    Muon Sys ($7.5$~kt)
    & 13\% & 33\% & 32\% 
    \\
    Nozzles $(0.14$~kt)
    & 79\% & 48\% & 48\% 
    \\
    \hline
    Total / bunch xs. 
    & 0.44 & 0.029 & 0.0053 
    \\
    Total / year
    & $1.5\times 10^{11}$ & $2.0\times 10^{10}$ & $1.5\times 10^{11}$ 
    \\
    \hline\hline
    \multicolumn4{c}{BIN exclusive reactions in HCAL and ECAL/year}
    \\
    \hline
    Total NC
    & $1.5\times 10^{9}$ & $4.6\times 10^{8}$ & $3.4\times 10^{9}$ 
    \\
    Total $\nu_e$ CC
    & $4.7\times 10^{9}$ & $1.4\times 10^{9}$ & $1.1\times 10^{10}$ 
    \\
    Total $\nu_\mu$ CC
    & $5.4\times 10^{9}$ & $1.7\times 10^{9}$ & $1.1\times 10^{10}$ 
    \\
    \hline
    ES $\nu_{e} e \to \nu_{e} e$ 
    & $2.0\times 10^{6}$ & $6.0\times 10^{5}$ & $7.3\times 10^{6}$     
    \\
    ES $\nu_{\mu} e \to \nu_{\mu} e$     
    & $3.8\times 10^{5}$ & $1.1\times 10^{5}$ & N/A     
    \\
    ES $\overline\nu_{e} e \to \overline\nu_{e} e$ 
    & $8.6\times 10^{5}$ & $2.5\times 10^{5}$ & N/A     
    \\
    ES $\overline\nu_{\mu} e \to \overline\nu_{\mu} e$     
    & $3.4\times 10^{5}$ & $9.9\times 10^{4}$ & $1.9\times 10^{6}$     
    \\\hline
    QE $\nu n \to \ell^- p^+$ 
    & $2.6\times 10^{6}$ & $2.5\times 10^{6}$ & $2.8\times 10^{7}$ 
    \\
    QE $\overline\nu p^+ \to \ell^+ n$ 
    & $2.7\times 10^{6}$ & $2.5\times 10^{6}$ & $3.2\times 10^{7}$ 
    \\
    Coh $\pi^0$ 
    & $3.0\times 10^{5}$ & $2.9\times 10^{5}$ & $3.5\times 10^{6}$ 
    \\
    Res $\overline\nu_{e} e \to \rho^-$     
    & $4.2\times 10^{5}$ & $7.7\times 10^{5}$ & N/A     
    \\
    Res $\overline\nu_{e} e \to K^{*-}$     
    & $2.6\times 10^{4}$ & $4.4\times 10^{4}$ & N/A     
    \\\hline
    IMD $\nu_\mu e \to \nu_e \mu^-$ 
    & $4.2\times 10^{6}$ & $1.2\times 10^{6}$ & N/A 
    \\  
    IMD $\overline\nu_e e \to \overline\nu_\mu \mu^-$ 
    & $1.2\times 10^{6}$ & $3.5\times 10^{5}$ & N/A 
    \\  
    ITD $\overline\nu_e e \to \overline\nu_\tau \tau^-$ 
    & $9.4\times 10^{3}$ & N/A & N/A 
    \\\hline
    Trident $e^+e^-$ 
    & $1.2\times 10^{6}$ & $2.9\times 10^{5}$ & $1.7\times 10^{6}$ 
    \\
    Trident $\mu^\pm e^\mp$ 
    & $2.9\times 10^{6}$ & $6.7\times 10^{5}$ & $5.0\times 10^{6}$ 
    \\
    Trident $\mu^+\mu^-$ 
    & $7.5\times 10^{5}$ & $1.6\times 10^{5}$ & $1.3\times 10^{6}$ 
    \\
    Trident $\tau^+\tau^-$ 
    & $6.1\times 10^{4}$ & $8.7\times 10^{3}$ & $5.0\times 10^{4}$ 
    \\\hline
\end{tabular}
  \end{ruledtabular}
  \caption{Total number of neutrino interactions in proposed muon collider detectors. The makeup of the event rate is shown as percentages for each detector component, excluding the magnetic solenoid.
  Rates for some exclusive scattering channels are also shown, including elastic scattering (ES), coherent $\pi^0$ production, resonant meson production on electrons, quasi-elastic (QE) neutrino scattering, inverse muon decay (IMD), inverse tau decay (ITD), and neutrino trident production on the Coulomb field of the nucleus.
  Channels with negligible rates due to the lack of $\overline{\nu}_e$ at $\mu$TRISTAN or due to kinematic thresholds are indicated by N/A.}
  \label{tab:event_rate_makeup}
\end{table}

\begin{figure}[t]
 \centering
  \includegraphics[width=0.465\textwidth]{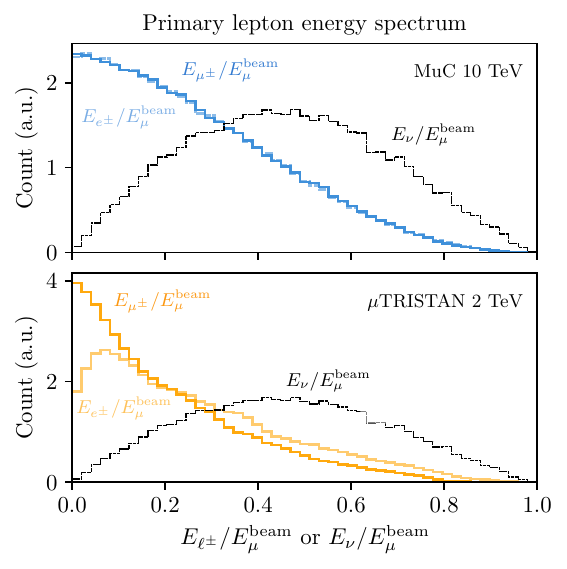}
  \includegraphics[width=0.49\textwidth]{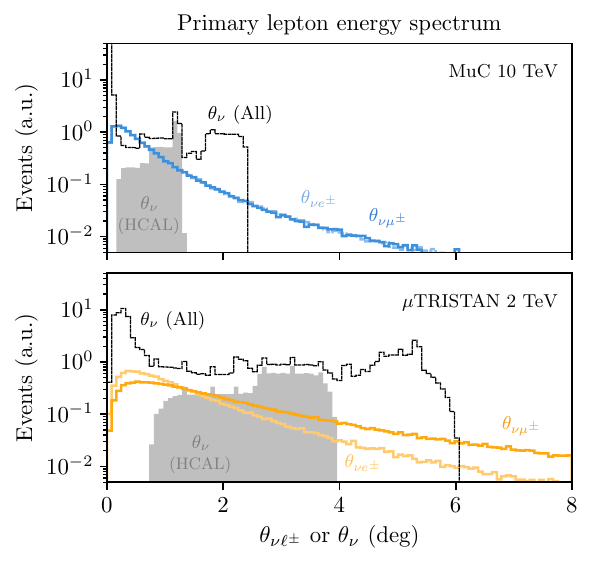}
 \caption{
 Top) the energy spectrum of charged leptons produced by BINs throughout the detector barrel for MuC 10 TeV and $\mu$TRISTAN benchmark designs.
 In dashed, we also show the energy of the BIN ($\nu$ or $\bar\nu$) that initiated the process.
 Bottom) The opening angle between the BINs and the outgoing lepton inside the detector for the same MuC benchmarks ($\theta_{\nu\ell^\pm}$).
 We also show the neutrino angle with respect to the detector axial direction in dashed black ($\theta_\nu$), isolating the contribution of HCAL events in solid gray.
 \label{fig:energy_spectrum}}
\end{figure}

In this appendix, we show more detailed properties of the BIN interactions.
\cref{fig:energy_spectrum} shows the energy spectrum of muons and electrons produced by BINs.
Different flavor components are shown in different shades.
The lepton energy spectra of $\nu_\mu$ and $\nu_e$ events are similar, but both are lower energy than their $\overline{\nu}_{\mu}$ and $\overline{\nu}_e$ counterparts due to the larger inelasticity of neutrino interactions. 
\Cref{fig:energy_spectrum} also shows the opening angle between the leptons and the parent neutrino ($\theta_{\nu\ell^\pm}$) as well as the angle of the neutrino with respect to the detector axial direction ($\theta_\nu$).
The structure seen in the latter is caused by the different amounts of material in the line of sight of the neutrino.
While most interactions happen in the highly forward direction, we are mostly interested in events with large $\theta_\nu$ as these can be better reconstructed at large detector radii.

At these energies, BINs are emitted mainly in the forward direction given the large muon boost, $\theta_\nu \sim 1/\gamma_\mu \sim 10^{-4}$, inheriting the parent muon beam properties at the time of production.
Conservation of angular momentum implies that the spectra of $\pbar{\nu}_e$ in $\mu^\pm$ decays are softer than $\pbar{\nu}_\mu$, resulting in a slightly higher interaction rate for the latter. Therefore, antineutrinos generally lead to harder final state leptons, albeit with a $\sim 2$ times smaller total cross-section. 
As the probability of backscatter is negligible at the TeV scale, the lepton angle is a good proxy of that of the neutrino.
In $60\%$ ($90\%$) of $\nu_\mu$CC events at the MuC-10-TeV, the opening angle between the neutrino and the outgoing muon is smaller than $0.6^\circ$ ($1.7^\circ$).
This correlation becomes tighter by requiring lower hadronic energy associated with the event.
The lepton charge can then be determined based on whether its momentum is tangential to $\mu^+$ or $\mu^-$ beam.

\emph{Beam wobbling---}
Neutrinos are also a radiation hazard at the locations where the beam exits the Earth's surface~\cite{Bevelacqua:2012odt,King:1999ja,King:1999kx}.
The proposed mitigation techniques, e.g., using bending magnets to induce a transverse wave-like trajectory for the muons~\cite{King:1999ptn,Skoufaris:2022wwg,Carli:2023wbr} (see also~\cite{Bandy:2009,King:1999kx,Geer:1997iz}), can be accounted for and should not pose a threat to a BIN program.
The proposed wobbling of the beam includes a change in the direction of the muons by no more than $\pm1$~mrad, with a sinusoidal longitudinal beam path and a new height that is at most $15$~cm larger~\cite{Carli:2023wbr}. While a more detailed study should take this wobbling into account, we can conclude that it affects the neutrino slice only by as much as $10$~cm in the most extreme cases.
Note that in the forward region, this wobbling can be significant, however, as it is designed precisely to spread the exit point of neutrinos from the Earth.

\emph{Beam polarization---} 
Muons produced from the decay in flight of pions are mildly polarized, with an average longitudinal polarization of $\mathcal{P_\mu}\sim 27\%$~\cite{Norum:1996mi,Palmer:1996gs,Blondel:2000vz}.
It remains an open question whether this polarization can be enhanced or maintained throughout the acceleration and storage stages in ionization cooling.
For $\mu$TRISTAN, on the other hand, muons from muonium are polarized by as much as $\mathcal{P_\mu}=0.9$~\cite{Hamada:2022mua} and the proposal aims to maintain $\mathcal{P_\mu}=0.8$ in the vicinity of the IP.
Because the energy distribution of BINs varies significantly with $\mathcal{P_\mu}$, the BIN interaction rate and energy spectrum are sensitive to it and $\mathcal{P}_\mu$ can be directly measured.
In the interval $0 < \mathcal{P_\mu}< 0.8$, the $\overline\nu_\mu$ ($\nu_e$) rate increases (decreases) by as much as $15\%$ ($25\%$); in addition, as shown in \cref{fig:polarization_ES}, the ES ratio in \cref{eq:Rthetaw} is also sensitive to $\mathcal{P_\mu}$.
This shows that a percent-level measurement of the $E_e$-dependent ratio in \cref{eq:Rthetaw} can constrain the residual polarization of the muon beam at the level of $1\%$ or better at the 10 TeV MuC.

\subsection{Exclusive interaction rates}
\label{app:exclusive_rates}

In this section, we show the makeup of the event rate in terms of exclusive scattering channels.
\cref{tab:event_rate_makeup} shows a list of such channels and the corresponding rate for each of our designs for a year of operation, including polarized and unpolarized versions of $\mu$TRISTAN.
The main scattering channels are deep-inelastic neutrino-nucleus CC and NC reactions.
These will contain a variety of hadronic final states.

Given the high rate of neutrino interactions, it is worth asking what type of rare exclusive processes could be accessed in the MuC detector.
Since the cross section for quasi-elastic (QE) and resonant scattering stops growing linearly with neutrino energy after about $\mathcal{O}($GeV) energies, the relative rate of these processes is roughly suppressed by GeV$/E_\nu$.
Here, we quote the QE rates as these can produce electromagnetic showers with a small hadronic energy activity, providing one type of background to neutrino-electron scattering events.

At a fraction of $\mathcal{O}(m_e/m_p)$, fully leptonic reactions appear, such as elastic and inelastic neutrino-electron scattering. 
As discussed in the main text, these channels are calculable in the SM and provide very clean channels with which to make precision measurements of electroweak parameters, constrain new physics, or measure beam parameters.
Other leptonic processes include inverse muon decay (IMD) $\nu_\mu e^- \to \mu^- \nu_e$, $\overline{\nu}_e e^- \to \overline{\nu}_\mu \mu^-$, and inverse tau decay (ITD), $\overline{\nu}_e e^- \to \overline{\nu}_\tau \tau^-$.
Both are only present in $\mu^-$ beams and have thresholds of $E_\nu^{\rm IMD} \sim 11$~GeV and $E_\nu^{\rm ITD} \simeq 3$~TeV.
In both cases, the charged lepton travels at extremely forward angles and appears without a hadronic vertex.
We also show the rate of resonant production of the $\rho$ and $K^{-*}$ mesons, $\overline{\nu}_e e \to \rho^- \to \pi^+ \pi^0$ and $\overline{\nu}_e e \to K^{-*} \to K^- \pi^0$, characterized by peaks at $E_{\overline{\nu}_e} \sim 600$~GeV and $E_{\overline{\nu}_e} \sim 800$~GeV, respectively~\cite{Brdar:2021hpy}.
These are also exclusive to the $\mu^-$ beam.

Other interesting semi-leptonic processes include neutrino-nucleus trident scattering: the production of pairs of charged leptons in the Coulomb field of the nucleus~\cite{Czyz:1964zz,Fujikawa:1971nx,Brown:1972vne,Altmannshofer:2014pba,Magill:2016hgc,Ballett:2018uuc,Zhou:2019frk,Zhou:2019vxt,Altmannshofer:2024hqd,Bigaran:2024zxk}.
At a MuC, the trident rate corresponds to a combination of $\nu_\mu + \bar\nu_e$ or $\bar\nu_\mu + \nu_e$ scattering rates.
In both cases, the final rate is a combination of CC and NC scattering of neutrinos off of virtual charged-leptons, representing a unique probe of flavored leptonic interactions and providing strong limits on new forces coupled to $L_\mu - L_\tau$, for instance~\cite{Altmannshofer:2014pba}.
The rates quoted in \cref{tab:event_rate_makeup} are obtained using a simplified rescaling of the oxygen and hydrogen results of \cite{Zhou:2019frk}.
Nuclear effects on the incoherent scattering components can be significant, but for most channels, coherent scattering dominates and provides the more promising detection channel.
With tens of millions of trident events per year at a MuC, this measurement would again be systematically limited and, with sufficient background rejection, would enable extraction of electroweak couplings and constraints on new physics coupled to the second and third lepton families.
This could improve on ongoing efforts to detect the $\tau$ production channels~\cite{Altmannshofer:2024hqd,Bigaran:2024zxk}.



\newpage
\appendix

\ifx \standalonesupplemental\undefined
\setcounter{page}{1}
\setcounter{figure}{0}
\setcounter{table}{0}
\setcounter{equation}{0}
\fi

\renewcommand{\thepage}{Supplementary Methods and Tables -- S\arabic{page}}
\renewcommand{\figurename}{SUPPL. FIG.}
\renewcommand{\tablename}{SUPPL. TABLE}

\renewcommand{\theequation}{A\arabic{equation}}

\onecolumngrid
\begin{center}
\textbf{\Large Supplementary Material}
\vspace{2ex}
\end{center}
\twocolumngrid

\section{Neutrino spectrum} 
\label{app:neutrinos_flux}

To obtain the neutrino flux, we sample the four-momenta of neutrinos produced in the decay of muons polarized either along or contrary to their direction of motion.\footnote{Here, we assume no transverse momentum is maintained in the muon beam. One can extend our work and method for transversely polarized muon beams and explore the corresponding physics.}
From the literature on neutrino factories~\cite{Cervera:2000kp,Broncano:2002hs}, we sample the differential muon decay rate neglecting the neutrino and electron masses ($(m_e/m_\mu)^2 \sim 2\times 10^{-5}$) but including radiative corrections.
For $\mu^-$ decays, the $\nu_\mu$ spectrum is given by
\begin{align}
\frac{\dd \Gamma}{\dd x \dd\cos \theta} &= \frac{G_F^2 m_\mu^5}{192 \pi^3}\left[ F_{\nu_\mu}^0(x) + \mathcal{P}_\mu J_{\nu_\mu}^0(x) \cos\theta \right.
  \\\nonumber &
\left. - \frac{\alpha}{2\pi}\left( F_{\nu_\mu}^1(x) + \mathcal{P}_\mu J_{\nu_{\mu}}^1(x)\cos\theta \right)\right].
\end{align}
where the functions $F^{0,1}$ and $J^{0,1}$ are given in Eqs.~19 to 22 of \cite{Broncano:2002hs} and $x = 2 E_{\nu}/m_\mu$.
For $\overline\nu_e$, one can just replace $\nu_\mu \to \nu_e$.
$\mathcal{P}_\mu$ is the ensemble average of the polarization of the $\mu^-$ or $\mu^+$ in the direction of motion ($\mathcal{P}_\mu = -1$ stands for $100\%$ left-handed $\mu^-$ and $\mathcal{P}_\mu = +1$ for $100\%$ right-handed polarized $\mu^-$).
For $\mu^+$, the same expressions hold with the replacement $\mathcal{P}_\mu \to - \mathcal{P}_\mu$.

\begin{figure}[t]
 \centering
 \includegraphics[width=0.49\textwidth]{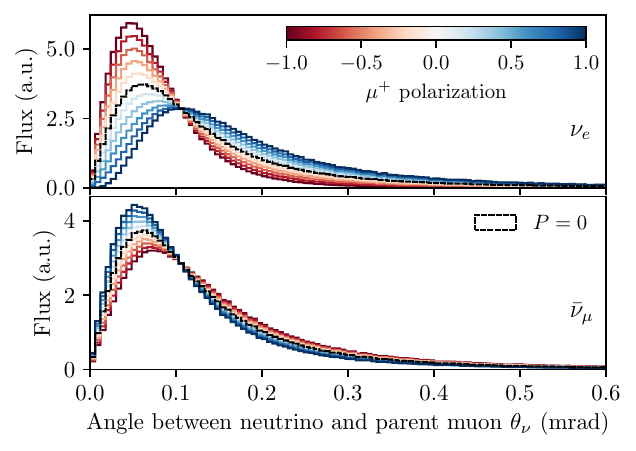}
 \includegraphics[width=0.49\textwidth]{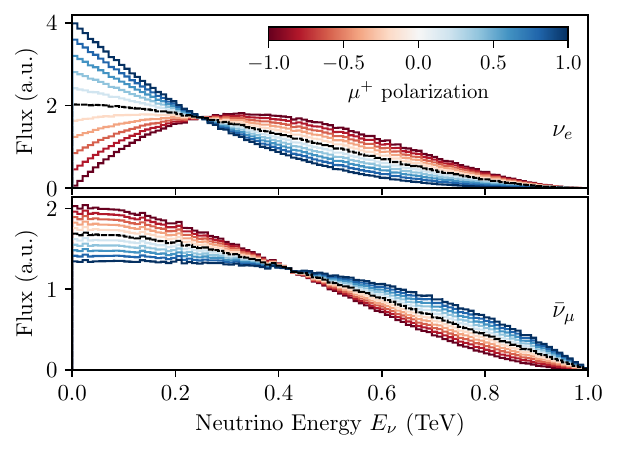}
 \caption{The distribution of angles between the neutrino and the parent muon (top) and neutrino energy (bottom) at $\mu$TRISTAN for varying degrees of $\mathcal{P}_\mu^+$ polarization.
 \label{fig:polarization_fluxes}
}
\end{figure}

\begin{figure*}[t]
 \centering
 \includegraphics[width=0.49\textwidth]{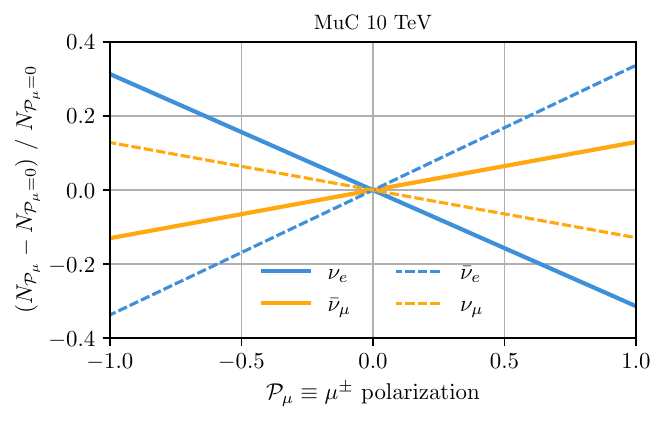}
 \includegraphics[width=0.49\textwidth]{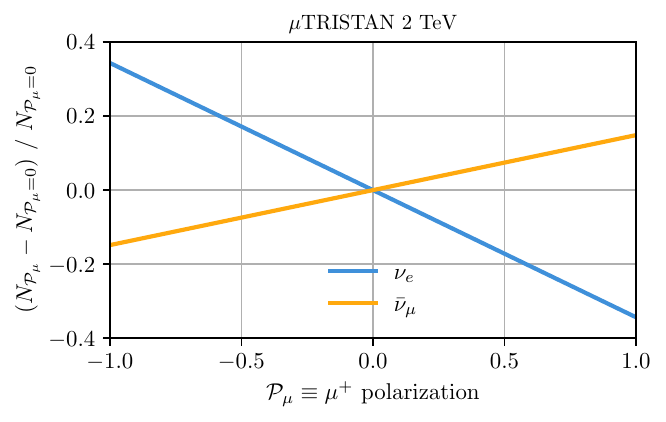}
 \caption{The relative variation in the total BIN interaction rate as a function of the muon beam polarization $\mathcal{P}_\mu$ for the $10$ TeV MuC (left) and the $2$ TeV $\mu$TRISTAN (right).
The polarization is assumed to be fixed throughout the entire ring.
 \label{fig:polarization}
}
\end{figure*}

The neutrino fluxes for a $1$~TeV $\mu^+$ beam are shown in \cref{fig:polarization_fluxes} as a function of the neutrino energy and the angle between the neutrino and the parent muon.
For $\mu^-$ decays, the fluxes are the same with the replacement $\nu_e \to \overline{\nu}_e$ and $\overline{\nu}_\mu \to \nu_\mu$.
The variation is most significant for outgoing $\nu_e$ thanks to angular momentum conservation.
As shown in \cref{fig:polarization}, the total neutrino interaction rate is very sensitive to the polarization of the beam.

\begin{figure*}[t]
 \centering
 \includegraphics[width=0.49\textwidth]{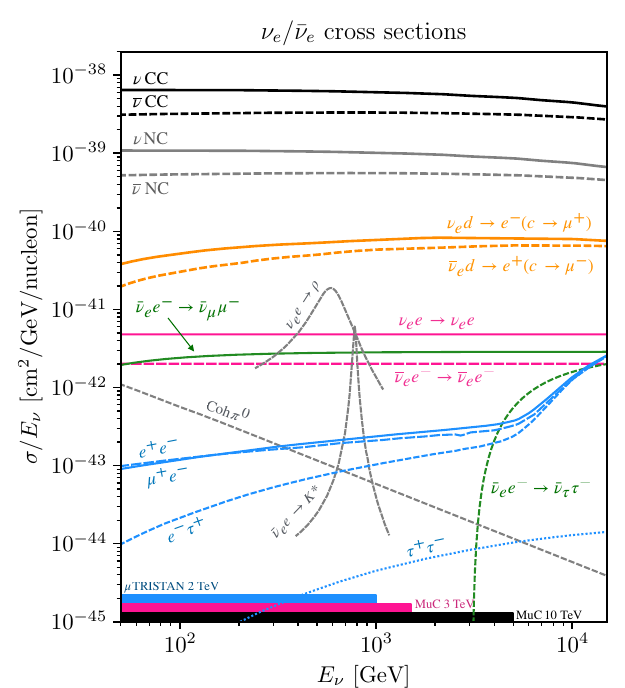}
 \includegraphics[width=0.49\textwidth]{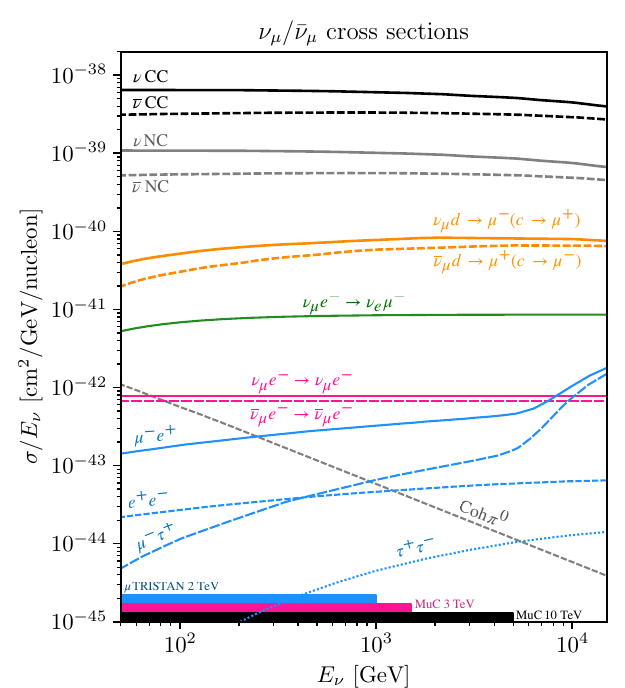}
 \caption{The neutrino cross sections used in this work normalized by nucleon number for $\nu_e/\bar\nu_e$ (left) and $\nu_\mu/\bar\nu_\mu$ (right). 
 In this example, we use neutrino-Carbon cross sections for coherent processes, and normalize all other processes to number of nucleons, neglecting nuclear effects.
 We show the charged-current (CC) and neutral-current (NC) total cross sections, as well as an estimate of the charm-induced dimuon production (see text for details).
 The energy range spanned by the neutrinos at different MuC benchmarks is shown at the bottom as a horizontal bar.
 \label{fig:nu_xsecs}
}
\end{figure*}

To obtain the rate of neutrino interactions, we track the momentum direction of daughter neutrinos and record those that intersect with the detector barrel.
The intersection line within each detector component is subdivided into smaller segments, and the probability of interaction in each component is calculated according to the density of the material (constant in each segment) and the total neutrino-nucleon cross-section.
Since the probability of interaction is much smaller than unity for each neutrino, the total interaction probability is given by the sum of the probabilities in each segment of constant density.
For each neutrino flavor and nuclear target pair, we then simulate the interaction products of the BIN using GENIE v3.04.00~\cite{Andreopoulos:2009rq, Andreopoulos:2015wxa}.
A compilation of the cross sections we use in this work can be found in \cref{fig:nu_xsecs}.

\begin{figure*}[t]
 \centering
 \includegraphics[width=0.5\textwidth]{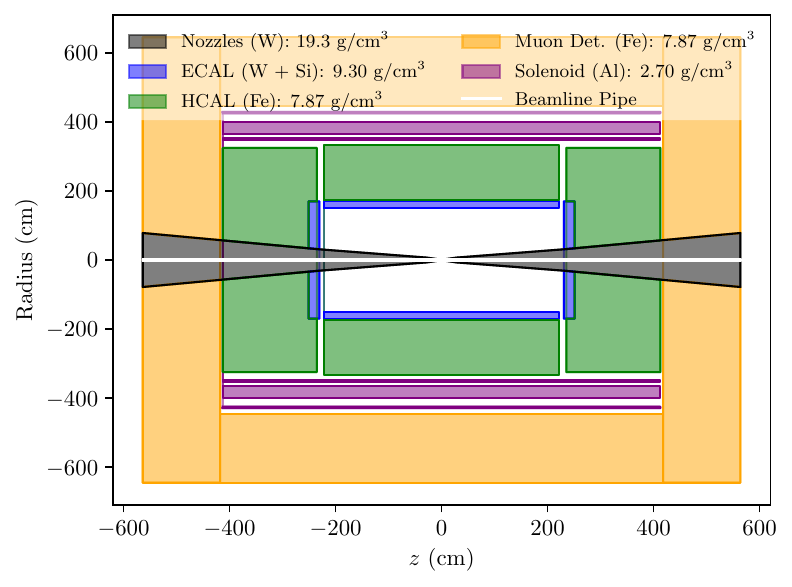}
 \includegraphics[width=0.41\textwidth]{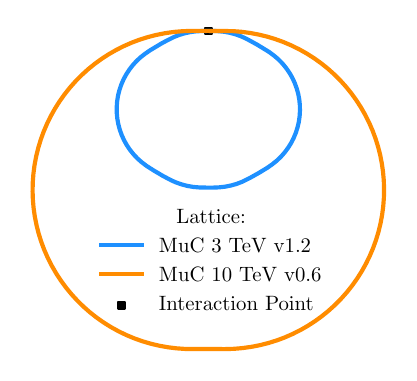}
 \caption{The detector used in this simulation, as per the dimensions in \cref{tab:detector_geometry}. Note that the tracker (outer, inner, and the vertex detector), composed of thin sheets of silicon, were not included in the simulation.
 \label{fig:detector_picture}
}
\end{figure*}

\begin{table*}[t]
 \centering
 \renewcommand{\arraystretch}{1.2}
 \begin{ruledtabular}
 \begin{tabular}{c|c|c|c|c}
Detector Parts & $\abs{z}$ (cm) & $R$ (cm) & Material &\ $N_{\rm targets}$ ($10^{24}$/$\mathrm{cm}^{-3}$)
  \\ 
  \hline\hline
Beampipe & $0 - 563.8$ & $0 - 2.2$ & Near vacuum & --\\
  \hline
Nozzles 1 & $6.5 - 230.7$ & $2.2, 2.2 - 31$ & W &$ 11.63$\\
  \hline
Nozzles 2 &\ \ $230.7 - 563.8$\ \ \ & $2.2 - 31, 2.2 - 78.2$ & W & $11.63$\\
  \hline
ECal (Barrel) & $0 - 221$ & $150 - 170.2$ & $0.38$W $+ 0.46$Cu $+ 0.1$Si & $6.99$\\
  \hline
ECal (Endcap) & $230.7 - 250.9$ &\ $31 - 170, 33.9 - 170$\ \ & $0.38$W $+ 0.46$Cu $+ 0.1$Si & $6.99$\\
  \hline
HCal (Barrel) & $0 - 221$ & $174 - 333$ &\ $0.75$Fe $+ 0.03$Al $+ 0.11$PS\ \ & $3.72$\\
  \hline
HCal (Endcap 1) & $235.4 - 250.9$ &\ $170 - 324.6$\ \ &\ $0.75$Fe $+ 0.03$Al $+ 0.11$PS\ \ & $3.72$\\
  \hline
HCal (Endcap 2) & $250.9 - 412.9$ &\ $33.9 - 324.6$, $56.8 - 324.6$\ \ &\ $0.75$Fe$ + 0.03$Al$ + 0.11$PS\ \ & $3.72$\\
  \hline
Solenoid (Inner) & $0 - 412.9$ & $348.3 - 352.3$ & Fe & $4.75$\\
  \hline
Solenoid (Middle) & $0 - 412.9$ & $364.9 - 399.3$ & Al & $1.63$\\
  \hline
Solenoid (Outer) & $0 - 412.9$ & $425 - 429$ & Fe & $4.75$\\
  \hline
Muon Detector (Barrel) & $0 - 417.9$ & $446.1 - 645$ & Fe & $4.75$\\
  \hline
Muon Detector (Endcap) & $417.9 - 563.8$ & $57.5 - 645$, $78.2 - 645$ & Fe & $4.75$\\
 \end{tabular}
 \end{ruledtabular}
 \caption{Dimensions and materials used for the detector simulation.As the detector is symmetric about $z = 0$, we only specify the absolute value of the axial coordinate $|z|$. The full detector layout can be obtained by reflecting these $z$ values onto $z < 0$. $N_{\rm targets}$ represents the nucleon density in each of the detector components. Double sets of values on $R$ in a single box indicate a conic section face that starts with the first dimensions and increases or decreases in height until the second dimensions.
 \label{tab:detector_geometry}
}
\end{table*}

\begin{figure*}[t]
 \centering
 \includegraphics[width=0.49\textwidth]{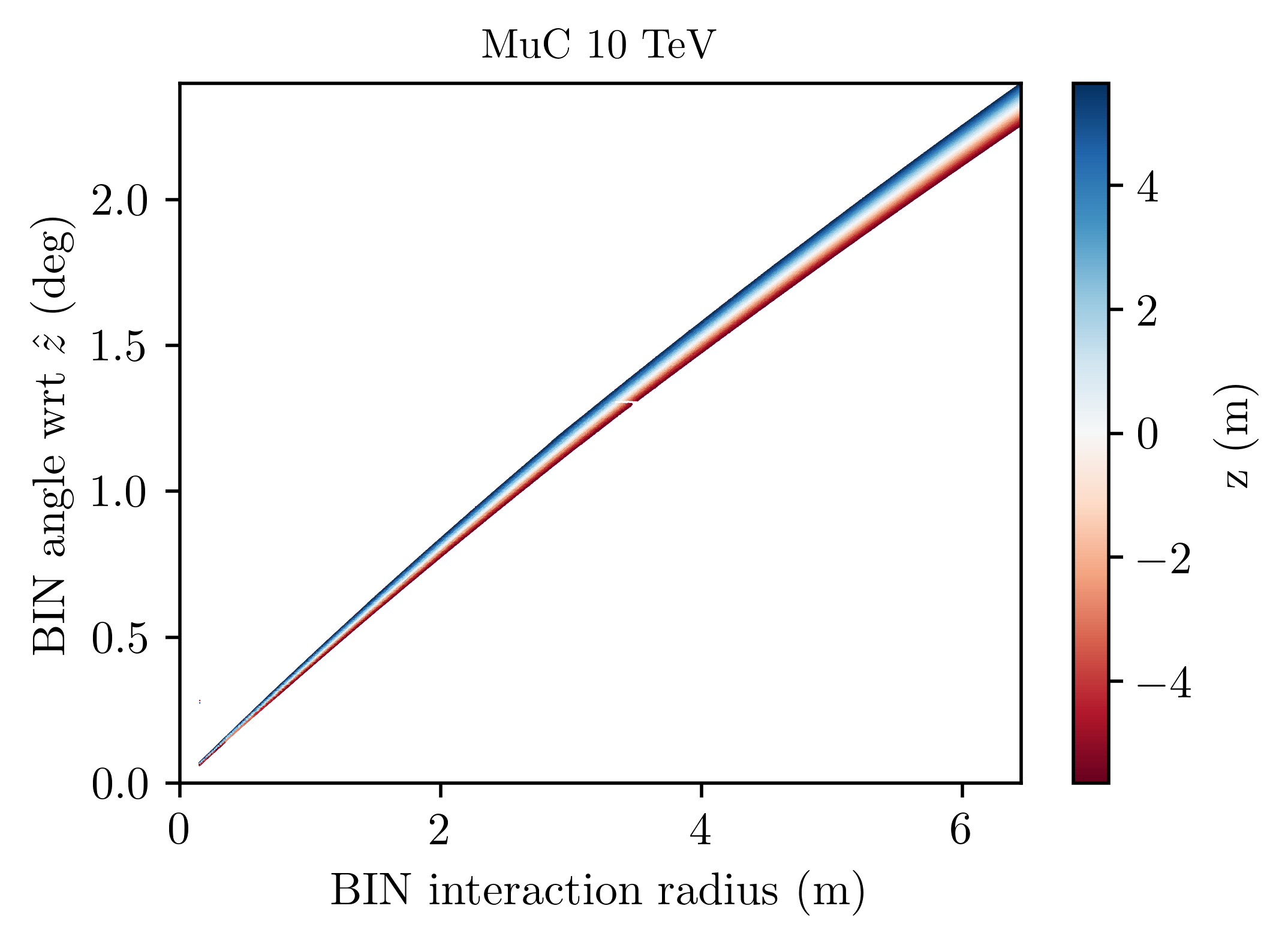}
 \includegraphics[width=0.49\textwidth]{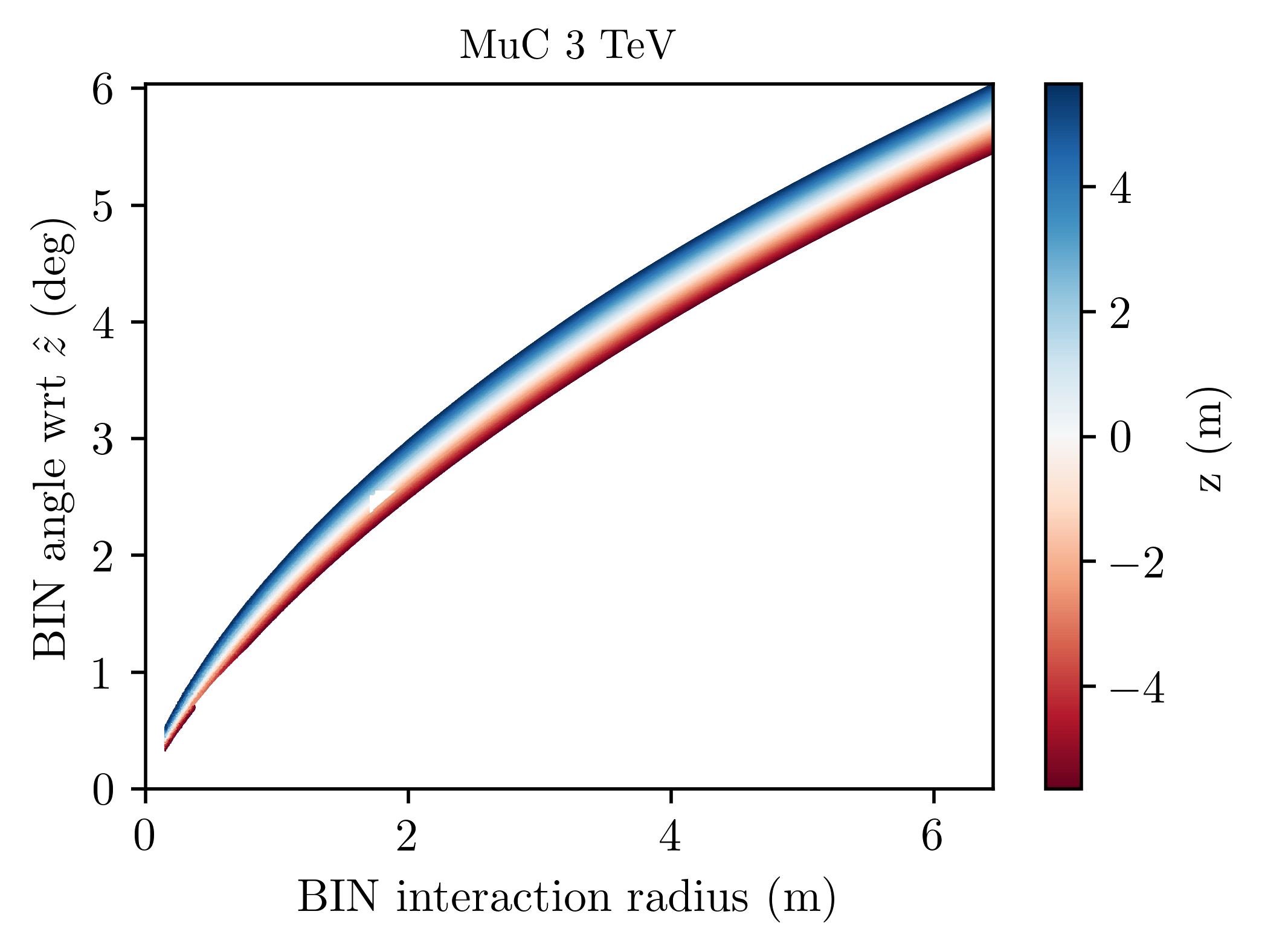}
 \caption{The neutrino angle wrt the detector axial position versus the radial position of the interaction for BIN events from one muon beam direction.
 The color indicates the $z$ position of the interaction within the detector.
 \label{fig:radial_correlations}}
\end{figure*}

\begin{figure}[t]
 \centering
 \includegraphics[width=0.49\textwidth]{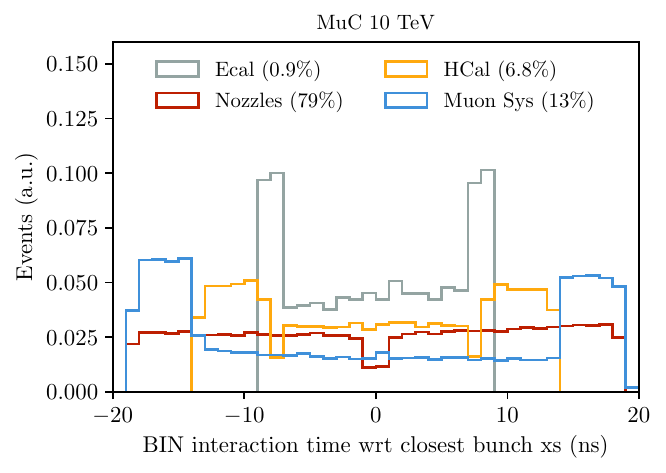}
 \caption{The time of BIN interactions in the detector with respect to the closest bunch crossing for the 10 TeV MuC benchmark.
 Each detector component is shown separately.
 \label{fig:timing}}
\end{figure}

\begin{figure*}[t]
 \centering
 \includegraphics[width=0.49\textwidth]{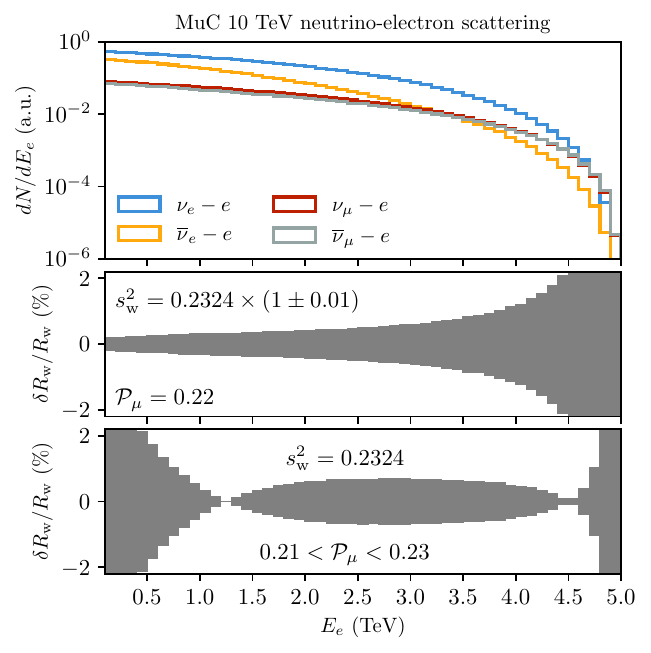}
 \includegraphics[width=0.49\textwidth]{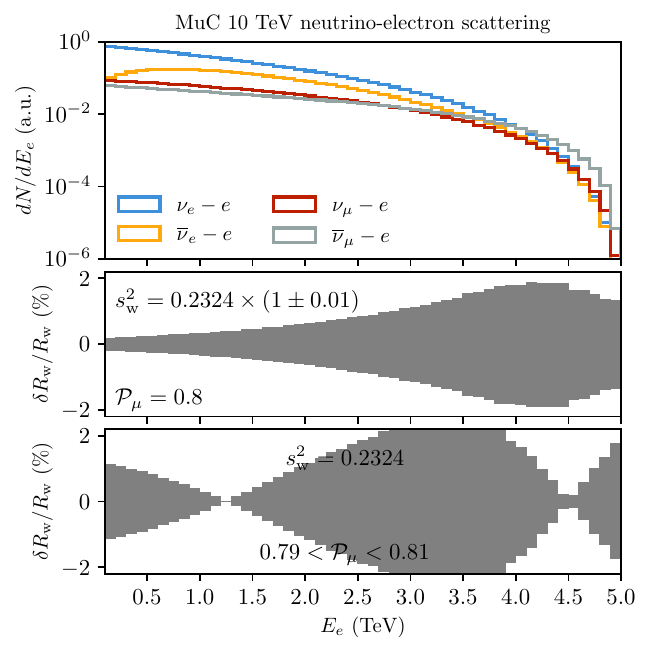}
 \caption{Top) The central value of the electron spectrum of ES events for different neutrino flavors at the MuC 10 TeV setting $\sinthetaw^2 = 0.2324$.
 Middle) the relative change in the ratio $\Rthetaw$ by varying $\sinthetaw^2$ by $1\%$.
 Bottom) the relative change when varying the beam polarization by $1\%$.
 On the left panel, we show the case for $\mathcal{P}_\mu = 0.22$ (left) and on the right, $\mathcal{P}_\mu = 0.8$.
 \label{fig:polarization_ES_others}
}
\end{figure*}

\section{Beam dynamics} 
We have adopted two approaches to simulate the muon beam and the collider ring: first, a simplified geometry made of a single straight section connected to a circular segment, and second, a more complete description of the accelerator lattice following the publicly available studies by the IMCC~\cite{MuCoL:2024oxj}.
In both cases, we assume the muon momentum to be $\sqrt{s}/2$, with $\sqrt{s}$ the center of mass energy of the primary muon collisions and add a Gaussian momentum spread of $0.1\%$ on the central magnitude of the momentum.
The neutrino angular spread with respect to TeV muons is of the order $\theta_\nu \sim 1/\gamma \sim 0.1/ (E_\mu/\text{ TeV})$~mrad and can be either comparable to or larger than the beam angular divergence.
We treat the divergence differently in the two accelerator models.

\paragraph{Simplified geometry:}
As a sanity check, we simulate the collider ring by modeling it as a straight section around the main detector joined on both ends by a large circular segment.
Together, they span a length of $C$ ($C = 8.7$~km for the MuC 10 TeV).
The straight (circular) segment is perfectly straight (circular).
We assume both muon beams travel in the same infinitely thin line to collide head-on.
In reality, collisions happen at a slight angle; for reference, the LHC crossing angles are a few hundred microradians, which is smaller than the assumed beam angular divergence.
We also neglect any directional modulation to the muon beam, such as those proposed to reduce neutrino radiation hazard.
The decaying muons are distributed evenly along the accelerator complex, using the interaction point as the injection point, which, within the limit of large boosts, is as good as any other location. 
In this simplified model, the angular divergence of the beam is modeled as a Gaussian with a constant standard deviation of $\delta \theta$.
For MuC we take $\delta \theta = 0.59$~mrad~\cite{InternationalMuonCollider:2024jyv} while for $\mu$TRISTAN we take $\delta \theta = 0.17$~mrad~\cite{Hamada:2022mua}.

\paragraph{IMCC lattice:}
Using the files provided by the IMCC~\cite{MuCoL:2024oxj}, we model the collider ring following the beam lattice and Twiss parameters provided. 
We force the trajectory of the beam center to follow the ring geometry exactly.
The ring shape is determined by a series of bending elements (dipole or multipole magnets), focusing elements (quadrupole, sextupoles, or multipoles), and drift (straight) sections.
In our implementation, bending elements are discretized into $\delta s=1$~cm straight elements for simplicity.
The geometric emittance of the beam is assumed to be a constant $\varepsilon = 0.5$~nm.
The transverse beam sizes in the plane of the ring ($\sigma_x$) and that transverse to it and the direction of motion ($\sigma_y$) are given by Gaussian with $\sigma_{x,y} = \sqrt{\varepsilon\beta_{x,y}(s)}$ where $\beta_{x,y}(s)$ is the Twiss beta function of the beam at the location $s$, provided by the lattice files.
Similarly, the angular divergence of the beam is given by $\delta \theta_{x,y} = \sqrt{\varepsilon\gamma(s)}$, where $\gamma(s) = (1+\alpha_{x,y}^2(s))/\beta_{x,y}(s)$, where $\gamma_{x,y}(s)$ and $\alpha_{x,y}(s)$ are the other Twiss parameters of the beam at the location $s$.
Our results are presented for the 10 TeV MuC use the lattice version \textsc{v0.6}, presented in more detail in \cite{Skoufaris:2023jnu}, and for the 3 TeV MuC and $\mu$TRISTAN, we use \textsc{v1.2}.
The latter is clearly a rough approximation for $\mu$TRISTAN as the beam energy is $50\%$ smaller than the 3 TeV MuC design.
Note that these designs do not include chicanes (dipole elements to sweep away secondaries in the beam) close to the interaction point.
The two ring shapes are shown in \cref{fig:detector_picture}.

As expected, the geometry of the accelerator is important and the total number of BIN interactions can vary by factors of $\mathcal{O}(2-5)$ depending on the exact choice of length of the straight section and the lattice.
In addition, the transverse momentum of the interaction products will depend on the curvature of the ring, as illustrated in \cref{fig:radial_correlations}.
The correlation between the interaction location (radius and $z$ position in the detector) and the incoming neutrino angle is strong, allowing to reconstruct the incoming neutrino direction on an event by event basis.

\cref{fig:timing} shows the timing of BIN events with respect to the closest bunch crossing.
The distributions follow the distribution of material in the line of sight of neutrinos and are separated into the individual detector components.
We do not subtract the distance between the BIN interaction vertex and the IP, the muon collision point.

\section{Detector details}
\label{app:detector_design}

Our benchmark detector design follows that of~\cite{MuonCollider:2022ded}.
The detailed geometry of all sub-components of the detector is included in \cref{tab:detector_geometry} and in \cref{fig:detector_picture}. 
These only represent one-half of the detector, which is assumed to be symmetric about the $z = 0$ plane. 
The nucleon density $N_{\rm targets}$ within the different parts are included to indicate the density differences.

\begin{figure*}[t]
 \centering
 \includegraphics[width=0.9
 \textwidth]{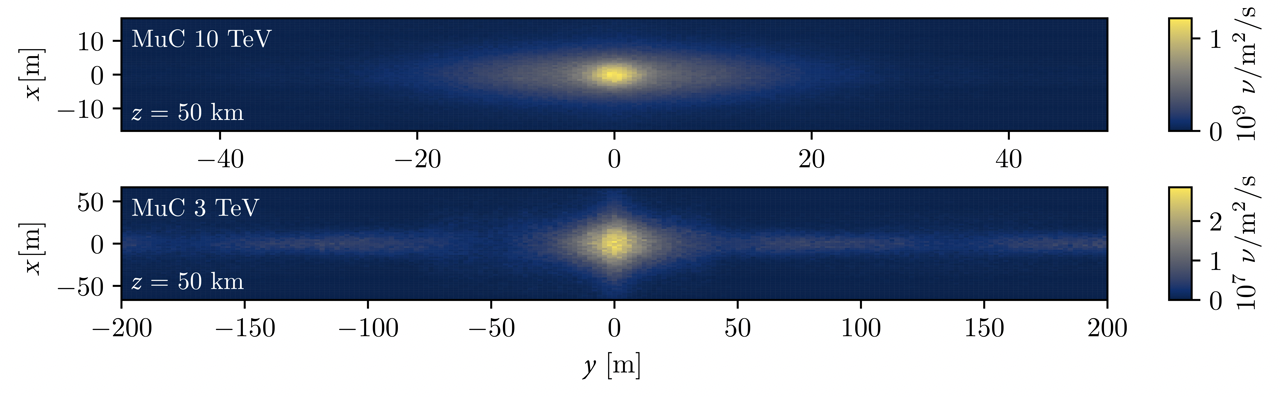}
 \caption{The neutrino flux at a distance of 50 km from the IP. The plane of the ring is aligned with the $y$ coordinate. The $x$ direction is normal to the plane of the ring.
 \label{fig:beamspot_flux}
}
\end{figure*}

\begin{figure*}[t]
 \centering
 \includegraphics[width=0.49\textwidth]{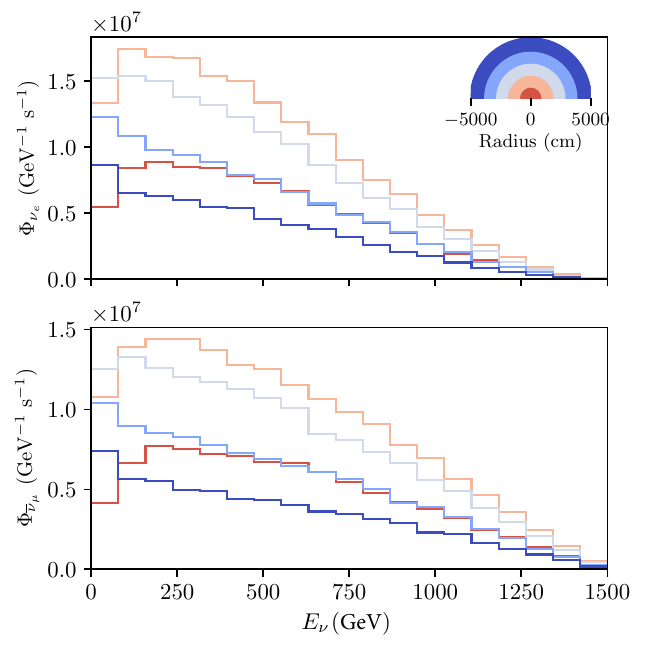}
 \includegraphics[width=0.49\textwidth]{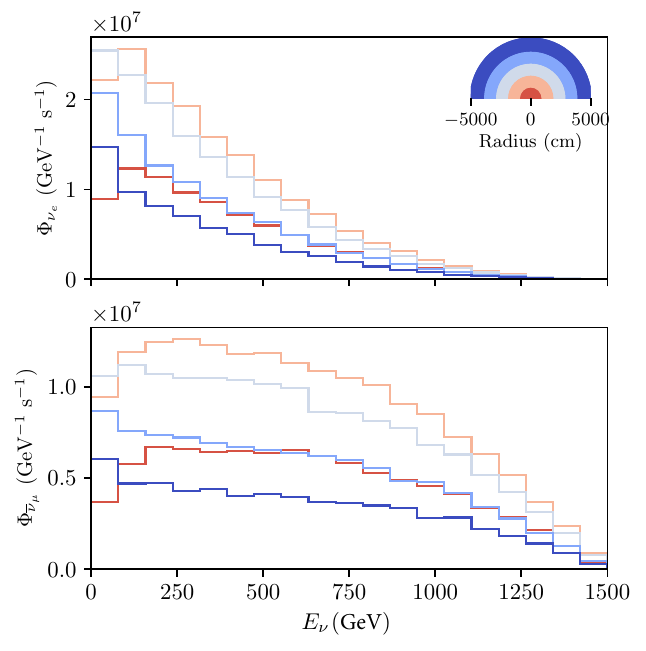}
 \caption{The number of neutrinos as a function of energy in various concentric rings. Each curve corresponds to a different mutually exclusive area, as their color indicates.
 \label{fig:forward_fluxes}
}
\end{figure*}

\section{Other Applications}
\label{app:other_applications}

In this section we discuss other applications of BINs. 

\emph{Deep Inelastic Scattering (DIS)---} Through DIS of neutrinos with nucleons, it is also possible to extract $s_w$ at large momentum transfer, $Q\sim 10$~GeV.~\footnote{For a complementary method to extract $s_w$ at $\mu$TRISTAN, see~\cite{Chen:2024tqh}.}
Most notably, this was done by the NuTeV experiment~\cite{NuTeV:2001whx}, leveraging the Paschos-Wolfenstein (PW) ratio~\cite{LlewellynSmith:1983tzz,Paschos:1972kj}, $R_{\rm PW} = \left(N^{\rm NC}_\nu - N^{\rm NC}_{\bar \nu}\right)/\left(N^{\rm CC}_\nu - N^{\rm CC}_{\bar \nu}\right)$, the ratio between the difference of neutrino and antineutrino NC and CC cross sections.
In the limit of an isoscalar target and a pure flavor neutrino beam, $R_{\rm PW} = 1/2 - s_w^2$.
The measurement was too large compared with the determination at LEP~\cite{ALEPH:2005ab}, but its status as an ``anomaly" is debated. 
Among possible solutions are a large strange sea-quark-momentum asymmetry in the nucleon~\cite{Barone:1999yv,Davidson:2001ji,NuTeV:2002ryj,Kretzer:2003wy,Ball:2009mk}, charge symmetry violation~\cite{Londergan:1998ai,Londergan:2003ij,Wang:2015msk}, and nuclear effects~\cite{Kovalenko:2002xe,Cloet:2009qs,Bentz:2009yy,Kulagin:2007ju}.
The PW ratio is not particularly useful at a muon source since the neutrino and antineutrino NC rates can only be differentiated statistically through energy distributions.
However, the well-known flux can provide a direct test of all the aforementioned hypotheses.
An enhancement in the sea-quark momentum asymmetry, for example, would be constrained by dimuon events from charm production $(\nu_\mu s \to \mu^- (c\to s \mu^+ \nu_\mu))$, improving on NuTeV's measurement~\cite{NuTeV:2007uwm}.

\emph{Standard candles---}
As discussed in the main text, neutrino-electron scattering provides a clean tool to constrain beam parameters and perform precision electroweak measurements.
In addition to \cref{fig:polarization_ES}, we also show the dependence of the ES ratio presented in \cref{eq:Rthetaw} for different beam polarization values in \cref{fig:polarization_ES_others}.
Clearly the shape of the electron recoil spectra varies signficantly for different beam polarizations, implying that the combination of a polarized and unpolarized beam can further constrain electroweak parameters, breaking any degeneracy between $\mathcal{P_\mu}$ and $\sinthetaw$, for example.
Other relevant rare processes at high energies include $\bar \nu_e e^- \to \ell^- \nu_\ell$, producing muons as well as tau leptons at thresholds of $11$~GeV and $3$~TeV, respectively.
Inverse muon decay produces extremely forward muons with $E_{\mu} \theta_{\mu}^2 \lesssim 2 m_e$, without hadronic activity.
These can be very effectively isolated from neutrino-nucleus reactions as demonstrated by CHARM-II~\cite{CHARM-II:1995xfh}, NuTeV~\cite{NuTeV:2001bgq}, and MINERvA~\cite{MINERvA:2021dhf}
(see also~\cite{Bogomilov:2013zba}).
Since these processes are exclusively present in the $\mu^-$ beam, they provide a clean channel with which to search for lepton-number-violating interactions such as
$\mu^+ \to \overline{\nu}_\mu e^+ \nu_e $ or $\nu_e e^- \to \overline{\nu}_\mu \mu^-$.

\emph{Sterile oscillations and non-unitarity---} New sterile neutrinos with $\Delta m^2_{41}$ of $\mathcal{O}(10^4 - 10^{5})$~eV$^2$ lead to BIN oscillations.
Through charge identification, $\bar\nu_\mu \to \bar\nu_\mu$ can be distinguished from $\nu_e \to \nu_\mu$ in the $\mu^+$ beam, and similarly for $\mu^-$.
For higher masses, the flavor transitions are independent of energy and, if $m_4 > m_\mu$, they will be non-unitary.
At a MuC, a constraint on the non-unitarity of the neutrino mixing matrix $U$ can be particularly interesting since the production is proportional to $G_F^2(1 - |U_{\mu 4}|^2) (1 - |U_{e 4}|^2)$ while the detection via $\nu_{e(\mu)}$CC is proportional to $G_F^2(1 - |U_{e(\mu)4}|^2)$, breaking the usual degeneracy that plagues conventional beams~\cite{Antusch:2009pm,Blennow:2016jkn}.
Because the beam has no $\nu_\tau/\overline{\nu}_\tau$ component, single-$\tau$ production would be a striking final state with which to constrain new physics.
This includes anomalous short-baseline oscillations $\nu_{\mu,e} \to \nu_\tau$, non-standard-interactions, and other exotic new physics.

\emph{Impact on forward physics---}
Double $\nu_\mu$CC BIN events can be a background to forward-muon physics, like measuring Higgs properties with $ZZ$ vector boson fusion~\cite{Forslund:2022xjq,Ruhdorfer:2023uea,Forslund:2023reu,Li:2024joa,Ruhdorfer:2024dgz}.
In both cases, the final state is a pair of high-rapidity muons traveling back-to-back.
The expectation for nozzle BIN events with 2 muons can be estimated by taking the Poissonian probability $P(\mu^+\mu^-) = \frac{\lambda^k e^{-\lambda}}{k!}$ with $k=2$ and $\lambda$ the rate of BIN muons per bunch crossing.
For instance, for the 10~TeV MuC, the interaction rate per bunch crossing in a single nozzle is about $\lambda \sim 0.1$ and $P(\mu^+\mu^-) = 0.0045$, and the total number of $\mu^+\mu^-$ events is around $3.8\times10^{8}$ events per year, excluding interactions in the beam pipe shielding and magnets.
With a luminosity benchmark of $1$~ab$^{-1}$/year at the 10 TeV MuC
and $\sigma(\mu^+\mu^- \to \mu^+ \mu^- h) = 8.7\times10^{-2}$~pb,
we expect $8.7\times10^{5}$ $ZZ$-fusion Higgses per year~\cite{Li:2024joa}.
The background rejection necessary to discriminate BIN events from these will depend on the forward muon tagging technique, although it is evident that BINs do not pose a significant threat to this program.
Firstly, the vast majority of BIN $\pbar{\nu}_\mu$CC events in the nozzles have pseudo-rapidity of $|\eta|> 6.5$, while $ZZ$ VBF muons can achieve excellent precision already for $|\eta| < 6$.
The transverse momentum of nozzle neutrino-induced muons is also much smaller, with the vast majority respecting $|p_T^\mu| < 10$~GeV at the IMCC.
Finally, the BIN-induced forward muons are more spread in time, and direction and do not connect to a track from the IP.

\emph{Light particle production---}New light particles produced in ultra-rare muon decays can also create a tangential flux of feebly-interacting states.
Examples of light states include heavy neutral leptons (HNLs), $\mu\to N\nu e$, flavor-violating axion-like-particles, $\mu \to e a$, and light bosons produced via bremsstrahlung, $\mu \to e \nu \nu \phi$.
These states will be highly boosted and can either scatter or decay as they pass through the kt-scale MuC detector.
As an example, let us consider a HNL that mixes with the muon or electron flavors.
In the long-lifetime regime, the probability for $N$ to decay into an $e^+e^-$ somewhere inside the kt-scale MuC detector is approximately given by $P \simeq \Gamma_N \ell / \beta \gamma$, where $\Gamma_N \sim (|U_{\mu 4}|^2 + |U_{e 4}|^2) G_F^2 m_N^5/192\pi^3$ and $\ell$ is the average distance $N$ spends inside the detector.
We assumed that, like the neutrinos, $N$ is tangential to the ring.
In that case, for the IMCC benchmark, the number of decay-in-flight events in a year is equal to
\begin{align}\label{eq:hnlrate}
  N_{N \to \nu e^+e^-} &\simeq \epsilon P_{\rm dec} \mathcal{B}(\mu \to N \nu e) 
  \\\nonumber &
  \simeq 100 \times \left(\frac{|U_{\mu4}|^2}{10^{-8}}\right)^2 \left(\frac{m_N}{30\text{ MeV}}\right)^6  \left(\frac{\epsilon}{3\times10^{-3}}\right),
\end{align}
where $\epsilon$ is the fraction of muons that decay with a HNL within detector acceptance.
Light particles need not only be produced in muon decays but can also arise from secondary interactions on the walls of the accelerator, for example.
A detailed study of light particle searches is left to future research.

\section{Forward Neutrino Flux}
\label{app:forward_flux}

The forward direction along the main straight section offers another opportunity to carry out neutrino measurements.
This direction will have the most concentrated neutrino flux in the collider ring, and even relatively small detectors could collect enormous statistics on neutrino interactions.
\cref{fig:beamspot_flux} shows the transverse spatial profile of the neutrino flux at a distance of 50 km from the IP for the 3 TeV and 10 TeV MuC benchmarks.
This corresponds to the distance between the IP and the surface of a spherical Earth with a ring located 200 meters underground.
\cref{fig:forward_fluxes} shows the energy of these neutrinos in various concentric rings centered around the IP at $z=50$~km for the 3 TeV MuC.
We also show an extreme case of polarization for MuC 3 TeV with $\mathcal{P}_\mu=0.8$.

Among the many interesting applications of such a facility is the precision measurement of fully leptonic interactions, such as $\nu-e$ elastic scattering.
Among the backgrounds to this measurement are $\pbar{\nu}_e$CCQE and coherent $\pi^0/\gamma$ production.
Both can be suppressed and/or constrained due to the differences in energy loss $dE/dx$ between electrons and photons, energy-angle correlations, and by requiring the absence of a hadronic vertex.
As opposed to conventional accelerator neutrino beams, half of the neutrinos in the beam are electron neutrinos, so suppressing low-hadronic activity $\nu_e$ interactions will be desirable.
In that regard, the $\mu^-$ beam is particularly advantageous for a \textit{magnetized gaseous detector}.
The detector magnetic field would allow the discrimination of positrons produced in $\bar\nu_e$CC interactions and the electrons in elastic scattering. 
In addition, photon backgrounds could be suppressed due to the lack of photon conversion in the gas.
With excellent vertex resolution, this should present an extremely clean environment for precise measurements for beam monitoring or fundamental physics.

One disadvantage of the forward neutrino flux measurement is the loss of information about the location of the parent muon decay through timing.
To a good approximation, the neutrinos from a given bunch arrive at the forward detector with the same longitudinal time structure as the muon bunch.
This is not true for the neutrino slice, where the timing and location of the event determines a baseline for the neutrino.

\bibliographystyle{apsrev4-1}
\bibliography{main}{}
\end{document}